\documentclass[runningheads]{llncs}
\usepackage{cite}
\PassOptionsToPackage{hyphens}{url}\usepackage{hyperref}

\usepackage{comment}
\usepackage[table]{xcolor}
\usepackage{todonotes}
\usepackage{amsmath,amssymb,amsfonts}
\usepackage{algorithmic}
\usepackage{graphicx}
\usepackage{textcomp}
\usepackage{courier}
\usepackage{bmpsize}
\usepackage{lipsum}
\usepackage{xspace}
\usepackage{todonotes}
\usepackage{hyperref}
\def\BibTeX{{\rm B\kern-.05em{\sc i\kern-.025em b}\kern-.08em
    T\kern-.1667em\lower.7ex\hbox{E}\kern-.125emX}}
\usepackage{soul}

\newcommand{\ledgerone}{Ledger One\xspace}
\newcommand{\ledgertwo}{Ledger Two\xspace}

\begin{document}

\title{Searching for a Trust Registry: Learning from Experiments with Hyperledger Indy}
\title{A Note on the Security and Scalability of Decentralized Identity: Learning from Experiments with Hyperledger Indy}
\title{A Note on the Blockchain Trilemma for Decentralized Identity: Learning from Experiments with Hyperledger Indy}

\author{Paul Dunphy}
\institute{OneSpan, Cambridge, UK\\
 \email{paul.dunphy@onespan.com}}

\maketitle

\begin{abstract}
The challenge to establish and verify human identity over the Internet in a secure and privacy-respecting way is long-standing. In this paper, we explore the blockchain trilemma of scalability, security, and decentralization in the context of the Trust Registry: a root of trust for a decentralized identity scheme that enables read and write access to shared records and is tamper-resistant. We make a case study of Hyperledger Indy -- an open-source technology bespoke for decentralized identity -- and conduct two empirical experiments to measure the latency of more than 45,000 transactions in the naturalistic environment of Amazon Web Services. We conclude that issues of Trust Registry scalability have multiple facets. While Hyperledger Indy captures data useful to underpin a decentralized identity scheme, the knock-on effect of its scalability limitations may indeed place constraints on properties of security and decentralization. The current credential verification process relies on transaction processing by a ledger with transaction processing bottlenecks, which may constrain the ideal of non-repudiation.
\end{abstract}

\section{Introduction}
The challenge to efficiently establish and verify human identity over the Internet is long-standing because the Internet does not innately have \textit{a way to know who and what you are connecting to}~\cite{Cameron2005}. Over time, while protocols such as TLS combined with trust anchored in web browsers became a ubiquitous approach to verify machine identity, there has been little consensus on how to digitally transact using a human identity that is flexible, secure, and respects privacy. For online services subject to anti-money laundering regulation, costs of verifying the identity of new customers in a process known as KYC (\textit{know your customer}) are estimated at £500m per year on average for a large financial institution; there are also time delays to establish an identity (a mean of 26 days)~\cite{ThomsonReuters2017}, and finally, user abandonment rates for identity establishment range between 65 – 95\%~\cite{AitGroup2018}. The landscape of tight (and changing) regulation coupled with inadequate identity technology incentivizes online services to establish and verify identity in invasive ways to users' privacy. Historical data breaches have surfaced unique identifiers from national identity schemes~\cite{Trak.In2020} and passports~\cite{CNET2018}; along with leaked personal information generally~\cite{CSO2020}. Furthermore, identity verification material has also been breached, such as images of passports~\cite{Wired2020}, national IDs~\cite{BusinessStandard2020}, and educational credentials~\cite{Trak.In2020}. 

Recently, there is a renewed interest to realize better methods of \textit{digital identity} from national governments~\cite{UKGovernment2021}, economic unions~\cite{EuropeanCommision2021}, and financial regulators ~\cite{FATF20}.Decentralized Identity (also known as Self-Sovereign identity (SSI)~\cite{Preukschat2021}) is one candidate approach to digital identity that is gaining traction. Decentralized Identity is \textit{user centric}~\cite{Bhargav-Spantzely2006} and is often defined by ten principles~\cite{Allen2016} that stand as a reinforcement to user-centric principles of privacy. Candidate schemes of this type have already been proposed~\cite{UPort2017,Maram2021,Dunphy2018}, and are distinguished by reliance on a \textit{Trust Registry}~\cite{ODonnell2019} - a root of trust for a decentralized identity scheme that provides tamper-resistant shared records that can be read from and securely written to by participants of the scheme. 

The \textit{Trust Registry} is a critical component of the vision for decentralized identity. However, there is no consensus on the optimal Trust Registry technology. Of over 80 publicly defined Decentralized Identifier (DID) methods (protocols to interact with Trust Registries) more than 95\% refer to \textit{algorithmic} trust registries~\cite{Smith2021} -- a category that includes distributed ledgers and blockchains. Furthermore, there is a lack of authoritative requirements to focus Trust Registry design on a given use case which poses a challenge since identity is highly contextual. The simultaneous lack of authoritative requirements juxtaposed to already existing Trust Registry implementations creates a pressing need to (i) explore and outline the characteristics of a successful Trust Registry for a given context; (ii) retrospectively understand the assumptions and requirements already embedded into Trust Registry implementations. The course of research is particularly pressing since researchers have highlighted the existence of a blockchain trilemma~\cite{EtherumWiki2022}, which takes that security, scalability, and decentralization are entangled and subject to a trade-off in implementation. The blockchain trilemma implies that Trust Registry implementations may not transfer to use cases with different expectations of security, decentralization, and scalability. As one example of the mismatch that can occur in the context of payments, research has already shown that the design of Bitcoin cannot securely support in-person payments in a typical mode of operation~\cite{Karame2012}. 

The contributions of this paper are the following:

\begin{itemize}
\item We propose Trust Registry design principles in light of the use case of \textit{know your customer} (KYC): a regulated process of identity establishment and verification that is governed by anti-money laundering laws. In doing so, we surface context-specific design decisions relevant to the blockchain trilemma of security, scalability, and decentralization \cite{EtherumWiki2022}.
\item We take Hyperledger Indy \cite{Hyperledger2019}  as a case study to explore how Trust Registry implementation choices might impact assumptions of scalability, security, and decentralization. Indy's consensus method,Indy Plenum 
\cite{Hyperledger2022}, is \textit{"special-purposed for use in an identity system"}. We conducted two measurement experiments where we measured the latency of more than 45,000 transactions in the naturalistic deployment environment of Amazon Web Services.[]We found that a more powerful client specification can reduce the latency of read requests. A larger network of Indy nodes leads to reduced capability to clear transaction backlogs. The storage of credential transparency and revocation mechanisms on the ledger and the need for all parties (including end-user client software) to read the ledger itself may impact its suitability for medium-sized federations and a sizable user group.
\item We note that one implicit design goal of Hyperledger Indy may be the security property of \textit{non-repudiation}. However, non-repudiation is a security property that is difficult to achieve in a system deployment. We discuss the implications for Trust Registry design.

\end{itemize}

\section{Related Work}

\subsection{User-centricity}
The ambition for users to exercise greater control and privacy over their personal information online extends more than a decade into the past. Establishing a prospective user/customer identity and verifying the associated evidence can be challenging, particularly where this process occurs over the Internet.  It is challenging to design standardized digital identity services and is likely impossible to build one universal scheme~\cite{Cameron2005}. Though, Microsoft made one high-profile attempt to create an identity meta-system for the Internet~\cite{Cameron2007} but withdrew support for it in 2011. However, there remains a pressing need for new and appropriately designed digital identity technologies~\cite{U.S.Government2010}, especially for services where verified identity information is essential to deliver a service.
Bhargav-Spantzel~\cite{Bhargav-Spantzely2006} describe an agenda for \textit{user-centricity} and discuss the specific aspects of enhanced \textit{control} in the context of identity.  
The term decentralized identity is not a technical term but appears to have a similar meaning to user-centricity. The term is present in a paper from 2007 by Weitzner~\cite{Weitzner2007} to capture the potential of OpenID to underpin URI-based identity, which he argues has the potential to provide the missing identity layer on the Internet. Fett et al.~\cite{fett2015spresso} argued that OpenID was not sufficiently decentralized nor privacy-respecting and, in the design of SPRESSO, proposed a single-sign-on scheme that leverages existing email addresses and ensures that a malicious identity provider is unable to track the relying parties where a user signs-on. 

\subsection{Know Your Customer (KYC)}
\label{sec:kyc}
Global anti-money laundering (AML) regulations cover how supervised organizations (e.g., financial institutions) must establish and verify the identity of customers. The specific measures fall under the umbrella of customer-due diligence, or more colloquially as know-your-customer (KYC). The Financial Action Task Force (FATF) is an organization convened by the G7 that periodically outlines principles to guide the design of KYC processes to influence legislation in member countries~\cite{FATF20}. Since the terrorist attacks on the United States in 2001, FATF has set a more aggressive global stance towards money laundering activities, resulting in stricter identity establishment and verification requirements. The KYC process is a point of tension between prospective customers and financial organizations. Institutions are subject to regulatory fines specific to their performance conducting KYC; for example, Deutsche Bank was fined £163 million in 2017 by a regulator in the UK~\cite{FinancialConductAuthority2017}. The KYC process itself is increasingly expensive to maintain, with some estimates that the largest institutions spent \$150 million on KYC in 2017, with an average of 26 days needed on average to complete one process~\cite{ThomsonReuters2017}.

\subsection{Blockchain-based Identity}
An identity can be thought of as a set of attributes associated with an entity~\cite{ISO2019} that apply to a given context. Research has started to explore the merits of using blockchains and distributed ledgers in identity schemes \cite{Dunphy2018w,Dunphy2018}. Specific scheme designs have used smart contracts to underpin cost-sharing schemes between identity providers and relying parties. Biryukov et al.~\cite{Biryukov2018} present a scheme leveraging smart contracts on the Ethereum blockchain. They modeled an onboarded user through a smart contract that securely stored a cryptographic accumulator. In combination with a witness, the user can evidence that their enrolment with an online service is active and not revoked. Parra-Moyano and Ross also describe the design of a smart contract-based solution~\cite{Parra-Moyano2017} that features a centralized database to store documentation provided by the end-user additionally. One limitation of their scheme is that it exhibits an over-reliance on a financial regulator's role to oversee the system's function. The part of smart contracts is primarily to capture payment between the verifier and the entity that establishes the user's identity.

Sonnino et al.~\cite{Sonnino2018} describe the design of a smart contract and identity credential framework that allows users to acquire and selectively disclose multi-show cryptographic credentials. They propose that four attributes of a credential framework are essential: distributed issuance, non-interactivity; blindness; unlinkability.

\subsection{Blockchain Scalability}
Scalability is a prominent theme in research on distributed ledger technology. Early work has focused on Bitcoin and how the time for blocks to settle creates risks to merchants of accepting Bitcoin \textit{faster payments}~\cite{Karame2012}. The scalability properties of the Bitcoin blockchain effectively rendered this use-case of in-person payments implausible. Croman et al.~\cite{Croman2016} provide scalability measures of Bitcoin and argue that parameter tuning of the network (e.g., block size) cannot overcome the fundamental scalability bottlenecks of Bitcoin in the network plane, e.g., that nodes must receive every transaction and perform validation locally before re-transmission to the peer-to-peer network.

An alternative route to scale blockchains is to leverage byzantine fault-tolerant protocols with coalitions of trusted parties. Vukolić~\cite{Vukolic2016} highlights the challenge to design systems that use byzantine fault tolerance as a consensus method and suggests how it can scale better through techniques that allow parallel computation of consensus.

Gorenflo et al. report on fine-tuning that can be performed on Hyperledger Fabric that can increase its throughput to 20,000 transactions per second~\cite{Gorenflo2019}. Notably, this was achieved without adjusting the Kafka-based consensus method but by adjustments to data handling in the transaction processing lifecycle.

\section{Searching for a Trust Registry}
We enumerate pillars of the design space that have a significant influence on the feasibility of a use case and are also given additional complextionby the use case of KYC (see Section \ref{sec:kyc}) and are subject to the blockchain trilemma of security, scalability, and decentralization. In doing so, we provide an extensible structure for thinking about context-specific Trust Registry design, enumerate constraints that can impact ideas for new solutions, and stimulate reflection on existing implementations. 

\subsection{Design Space}
The elements listed below are not necessarily mutually exclusive, but their categorization highlights the structure of the design space.

\subsubsection{Determine Stability of State}
A Trust Registry with an actionable state can support decisive actions in the real world and provide evidence to resolve disputes. Public-permissionless blockchains (e.g., Bitcoin ~\cite{Nakamoto2008}, Ethereum \cite{Wood2014}) rely upon Nakamoto consensus, which provides probabilistic finality on the ledger state. However, while probabilistic finality appears acceptable as a way to determine ownership of Bitcoin, it may not be sufficient in KYC if a service must base an identity verification decision on the shared records at a specific time  $t$, only for that state to be usurped by a latent update from $t-1$ (e.g., a revocation). Online services subject to KYC are likely to retain an adversarial position towards each other since they are competitive organizations in the same market, which means that \textit{finality of state} may be desirable to justify an action or to resolve a dispute. However, while the judicious choice of consensus method can assure real-time ledger consistency, it can also introduce scalability limits on a network.

\subsubsection{Define and Ensure Ongoing Fairness}
The property of decentralization is one way to evidence an identity network's fairness. However, while a Trust Registry can be decentralized by design, it may be less decentralized in practice. One inherent challenge facing an identity network that features the roles of an identity provider (IdP) and a relying party (Rp) is to account for the asymmetry of effort. The IdP invests time to bootstrap an identity for the first time (a process widely associated with high costs in the financial sector \cite{ThomsonReuters2017}), and the \textit{Rp} incurs a lesser cost to verify identity. Therefore designers can consider mitigating \textit{free riders} \cite{Cila2020} -- where the relying party generally avoids taking the role of IdP in a transaction. Prior work aims to address cost imbalances by leveraging cryptocurrency as compensation for the identity provider ~\cite{Biryukov2018, Parra-Moyano2017}. However, this is just one specific problem of fairness. Suppose a subset of stakeholders are routinely \textit{out-voted} in consensus protocols by coalitions of organizations with shared behavior. Technical mechanisms to enforce greater decentralization can constrain scalability. Therefore, governance frameworks are an essential means to ensure ongoing fairness (e.g., Sovrin Foundation \cite{Sovrin2016}) and to define penalties for non-compliance. 

\subsubsection{Choose an Adversarial Stance}
Any Trust Registry must protect itself against realistic adversarial scenarios to maintain trust in the system but must avoid over-design of security lest there be a reduction in decentralization and scalability.  Attacks can take many forms, such as deliberately sending erroneous or latent updates to the Trust Registry or involuntary attacks through client software bugs. Designers must consider how confidence in the Trust Registry can persist in the face of a security incident -- since security incidents are inevitable.  If the Trust Registry should remain functional in the face of a threat to its integrity, then Byzantine fault tolerance \cite{Lamport1982} is a technique of reaching consensus on the state in the presence of faults. Confidentiality threats could arise through an honest-but-curious attacker passively searching for insights about other organizations or end-users. Threats to the availability of the Trust Registry might lead to the creation of an access control mechanism, but this could limit applications that depend upon end-users accessing the ledger state.

\subsubsection{Provide Privacy for Organizations}
A decentralized identity scheme in the financial sector is likely composed of parties in direct competition in a market. Therefore, a Trust Registry should support confidentiality by default and privacy mechanisms to enable minimal disclosure~\cite{Cameron2005}. An organization may have to disclose information to several entities in a decentralized identity network, for example, end-users through GDPR laws, an IdP or Rp to resolve a dispute, or even to a reputable auditor or regulator of the system to evidence conduct. Furthermore, we should consider how privileged insights can be derived from the data structures on the Trust Registry or information flows received in peer-to-peer messages. Any unanticipated disclosures can form a reason to abandon an identity network. Therefore, techniques such as ring signatures \cite{Clack2019} could be helpful to authenticate messages from a trusted group without precisely knowing the signer, or zero-knowledge proofs~\cite{Feige1988} that can provide unlinkable disclosures that do not leak information.

\subsubsection{Who Audits, and How?}
Prior work has assumed that regulators require transparency of the data generated by decentralized identity schemes in the financial sector to mitigate risks to the system overall~\cite{Parra-Moyano2017}. However, while regulators of the financial industry may require approval of a specific decentralized identity scheme (as implied by the FATF~\cite{FATF20}), they may not need a live global view of Trust Registry data to oversee identity schemes. Other pre-existing regulations might be more relevant than new regulatory actions or technical vantage points. For example, in the UK, financial institutions already provide transparency on matters of fraud to regulators through the submission of an Annual Financial Crime Report \cite{FinancialConductAuthority2022}; this report comprises fraud statistics and an enumeration of risks to the institution and the controls in place to mitigate those risks. On the specific topic of new identity schemes, the UK government envisions a back-seat role to facilitate the creation of a national trust framework \cite{UKGovernment2021} to set the overall scope and ground rules of new schemes. Therefore, designers may wish to weigh the impact of assigning new roles and technical oversight to regulators in a new system where the political, cultural, or practical precedent does not exist. 

\subsubsection{Anticipate Transaction Patterns}
\label{sec:deploy}
There is an implicit yet prevailing assumption that a decentralized identity scheme's scalability is lower than that needed for financial transactions. However, underestimating the transaction activity of a scheme will likely create new constraints on security and decentralization. One helpful heuristic to evolve a model of usage is the phenomenon of \textit{The Big 4}, which depicts a common market scenario where four organizations obtain a dominant position in an industry. Taking the banking example in the UK context, there are 65 million checking accounts in total, and the big 4 possess 77\% of those accounts.~\cite{CompetitionandMarketsAuthority2014}. This model's simple extension and analysis can highlight various requirements, such as the need to bootstrap a decentralized identity scheme with 50 million credentials on day one. Furthermore, the numbers invite us to envision how the maintenance of those 50 million credentials also creates a transaction load on the network (keeping in mind that the operations of reading and writing each generate one transaction to be processed).  Finally, there are also different implications depending on who can create transactions; for example, is this privilege restricted to online services themselves, or can end-users interact peer-to-peer with the Trust Registry too?

\section{Case Study: Hyperledger Indy}

\begin{figure}[t]
\centering
\includegraphics[width=0.9\textwidth]{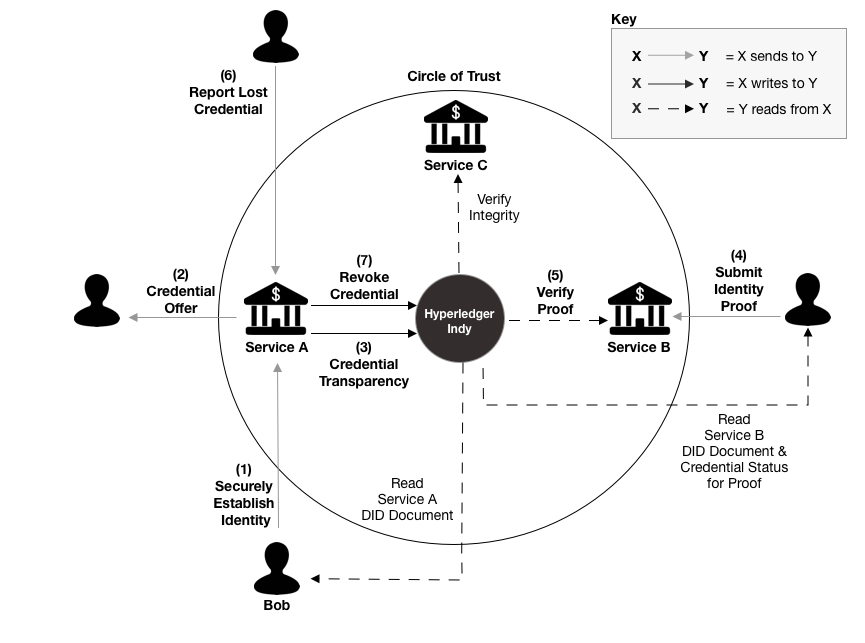}
\caption{Overview of the key actors and operations in a Hyperledger Indy workflow and the key steps in the credential lifecycle. In this example, Service A is the identity provider, Service B is a relying party, and Service C is another stakeholder within the circle of trust that is not involved in the interaction with Bob (the user) but has a role in responding to read requests for ledger data.}
 \label{fig:actors} 
\end{figure}

\begin{table}[t]
\small
\begin{center}
\begin{tabular} {  p{5.5 cm} |  p{6.5cm} }

Principle & Evaluation  \\
\hline
Determine Stability of State & Consensus method is Indy Plenum~\cite{Hyperledger2022} which achieves settlement finality at the end of each round of consensus. History is not expected to change. \\
Define and Ensure Ongoing Fairness & One organization is likely to administer one Indy node, and each node has one vote in the consensus process. No inherent mechanism for addressing cost imbalance.  \\
Choose an Adversarial Stance & Byzantine fault tolerance that assumes the majority of network nodes are honest to confirm write transactions. By default, at least two nodes should corroborate the content of a read transaction. No access control. \\
Provide Privacy for Organizations & No confidential information is stored in the ledger. Publicly declared Decentralized Identifiers (DIDs) and public keys are viewable along with credential definitions and schemas. Enumeration of issued credentials is not possible due to the use of cryptographic accumulators~\cite{stanislaw_jarecki_accumulator_2009}. \\
Who Audits, and How? & No explicit approach to permit system audit. Assumed process of self-assessment and self-disclosure. \\
Anticipate Transaction Patterns & No explicit transaction processing goals available relative to a use case. \\

\end{tabular}
\caption{Evaluation of Hyperledger Indy according to pillars of the design space.}
\label{table:indyrationale} 
\end{center}
\end{table}

Hyperledger Indy~\cite{Hyperledger2019} (herein Indy) is one candidate to underpin a Trust Registry. Indy is interesting for several reasons. Firstly, Indy is purpose-built for decentralized identity and includes a consensus protocol known as Indy Plenum, purportedly designed for identity applications and based upon a redundant Byzantine fault-tolerant state machine replication protocol~\cite{pierre-louis_aublin_rbft:_2013}.  Secondly, it underpins a public distributed ledger deployment by the Sovrin foundation~\cite{TheSovrinFoundation2016}. Thirdly, Indy is open-source. Table \ref{table:indyrationale} illustrates a qualitative evaluation of Indy in relation to Trust Registry design parameters. Figure \ref{fig:actors} illustrates an abstraction of the Hyperledger Indy identity lifecycle. As with other decentralized identity scenarios, the three protagonists are the identity provider (Service A), relying party (Service B), and end-user (Bob). These three entities can also be equivalently referred to as issuer, verifier, holder~\cite{Preukschat2021}). The operations in the figure are implemented in the form of multiple SDK calls illustrated in Table \ref{table:transtoproc}.

Decentralized Identifiers (DIDs)~\cite{Sporny2021} are an emerging type of identifier that can be \textit{"decoupled from centralized registries, identity providers, and certificate authorities"}. DIDs are implemented in Indy though the term \textit{"NYM"} is predominantly used throughout the SDK since presumably it pre-dates DID. Many decentralized identity interactions are likely to be entirely confidential. However, some may be bootstrapped using publicly advertised DIDs securely mapped to a DID Document: the meta-data of a DID, e.g., public key, that is tamper-evident and securely mapped by a Trust Registry. 

An identity provider must establish the identity of a user according to the norms of the domain (e.g., passport, driver's license) and, at the end of the process, issue the user with a \textit{credential}: a cryptographic attestation that expresses confidence in the veracity of specific user attributes. A credential can be thought of as a simple certificate configured to contain human attributes where the creator of the attestations can be authenticaticated. The implementation of credentials is realized through IBM's Idemix~\cite{noauthor_specification_2010} technology and represented in a format that is currently standardized with the W3C~\cite{W3C2019}.  The credential operations themselves require several on-ledger data items. The type meta-data for a credential is stored on the Indy ledger as a credential definition (referenced as CRED\_DEF); this is a unique identifier that references the data fields that credentials of this type should contain, also a pointer to the credential \textit{revocation registry} (REVOC\_REG), which references the cryptographic accumulator for credential issuance transparency and revocation checking~\cite{stanislaw_jarecki_accumulator_2009}. An identity provider issues a new credential through the call (REVOC\_ADD) is made to the ledger, and when a credential is revoked, the function (REVOC\_DEL) is called. Each revocation registry contains a pointer to the revocation registry definition (REVOC\_DEF).

A \textit{relying party} provides an online service to end-users and is capable of verifying the identity of the user through verifiable data such as a credential. To receive a service from the relying party, the end-user must build an identity \textit{proof} using software on a personal device. An identity proof could be as simple as a presentation of an eponymous credential along with proof of ownership of that credential (e.g., through a link secret~\cite{Hyperledger}), or something more complex such as a zero-knowledge proof~\cite{Feige1988}.

Finally, when building a proof, a user must read the ledger to calculate the value of the cryptographic accumulator up to a given point in time to prove that their credential was valid in a given time window (REVOC\_DELTA).

\begin{table*}[t]
\small
\begin{center}
\begin{tabular} {| l | l |c | c | c | c | c | c | c |}
\hline
Shorthand & Indy SDK request (node.js) & (1) & (2) & (3) & (4) & (5) & (6) & (7) \\
\hline
\texttt{\detokenize{READ_DID}}                             &    buildGetNymRequest(...)            & U   & -       & -      &  U     & -     & -        & -          \\
\texttt{\detokenize{READ_CRED_DEF}}                  &     buildGetCredDefRequest(...)     & -    & U   &  -     &  U   & RP       &  -   & -       \\
\texttt{\detokenize{READ_REVOC_DEF}}      &     buildRevocRegDefRequest(...)  &-    &-         &-      &  U   & RP       &   -  & -        \\
\texttt{\detokenize{READ_REVOC_REG}}                &    buildGetRevocRegRequest(...)  & -   &-          & IdP &  U     & RP      & -   & IdP        \\
\texttt{\detokenize{READ_REVOC_DELTA}}   &     buildGetRevocRegDeltaRequest(...)  & -   &-         & -     &   U     & RP      & -   & -       \\
\texttt{\detokenize{WRITE_DID}}                              &    buildNymRequest(...)                  & -   &-         & -     & -      & -     & -      & -             \\
\texttt{\detokenize{WRITE_REVOC_ADD}}  &    buildRevocRegEntryRequest(...)  & -   & -        & IdP & -       & -    & -      & -             \\
\texttt{\detokenize{WRITE_REVOC_DEL}}   &    buildRevocRegEntryRequest(...)  & -   &-          & -      &  -   & -  & -        &  IdP    \\
\hline
\end{tabular}
\caption{The mapping between steps of the model in Figure \ref{fig:actors} and the Hyperledger Indy-specific operations that are needed to accomplish those steps. Note that one conceptually simple operation in the model can require multiple operations with the ledger to be issued in Hyperledger Indy. The numbers along the horizontal axis of the table correspond to: (1) Establish Identity; (2) Credential Offer; (3) Credential \& Transparency; (4) Identity Proof; (5) Verify Proof; (6) Report Lost Credential; (7) Revoke Credential.}
\label{table:transtoproc} 
\end{center}
\end{table*}

\section{Experiments}
To learn more about the blockchain trilemma in the context of Hyperledger Indy, we designed experiments to capture empirical data to evidence the scalability of the different operations of the Hyperledger Indy distributed ledger and quantitatively inform our exploration of the blockchain trilemma in decentralized identity. 

\subsection{Methodology}
The quantitative approach of contrasting transaction processing under different scenarios is a common way to illustrate the performance of blockchain architectures and inform discussions about system acceptance \cite{Danezis2015}.  Therefore, we wanted to mirror this approach and investigate transaction processing performance in an optimal case for Hyperledger Indy and in more challenging conditions under ambient transaction loads.  

We conduct two experiments: (i) Baseline Experiment; (ii) Transaction Load Experiment. The independent variable across both experiments is the number of ledger nodes participating in a Hyperledger Indy network: \ledgerone has four nodes; \ledgertwo has 11 nodes. The dependent variable across both experiments is \textit{transaction latency}, which captures the return trip of a transaction between sender and receiver. A \textit{read request} is a message to a ledger node to lookup the value of a data item recorded in the ledger. We refer to the transaction latency of read requests as \textit{read latency}. A {write request} is a digitally signed message indicating a change to the state of a data item on the ledger. We refer to the transaction latency of write requests as \textit{write latency}.

Transaction latency measurements are difficult to generalize with confidence. However, if experiments are transparently described and leverage widely available technology, there can be much to learn about specific scenarios of deployment, e.g. best and worst cases.

\subsection{Experiment Architecture}
\label{sec:arch}
We used the tools and infrastructure provided by Amazon Web Services (AWS) to conduct our experiments. We created three virtual machine (VM) templates: Ledger Node, Measurement Client, and the Load Generator.

Each Hyperledger Indy ledger comprises multiple Ledger Node instances running the following software: Ubuntu 16.04 6 operating system; Node.js 12.13.0; Indy-sdk 1.11.1-dev-1318; Libzmq 4.1.4-7ubuntu0.1\footnote{We carried out this work in Autumn 2020, a time when these software versions were the most current.}. Each VM is a t2.xlarge Amazon VM.

The Measurement Client initiates and measures the read and write requests under measurement. When the measurement node sends a request, the node waits for a response from an appropriate number of ledger nodes and records the latency. 

The Load Generator controls a burst of ambient transactions directed towards ledger nodes. The Load Generator (only used in Experiment Two) has a target for the rate of transactions to be sent (e.g., 20 transactions per second) and sends requests either faster or slower to maintain that rate, and does not process responses from a Ledger Node.

\subsection{Experiment 1: Baseline Experiment}

\subsubsection{Method}
The experiment has a 2X2 design (Ledger $X$ VM). We included two additional independent variables in this experiment, reflecting different measurement client specifications: t2.micro and t2.medium (in AWS parlance). We configured the Measurement Client to send 1000 transactions for each of the eight requests in Table \ref{table:transtoproc} which means we measured 32,000 transactions in total.

\subsubsection{Results}
Table \ref{tab:baselineResults} provides an overview of the transaction latency results. To determine any overall effect of a specific ledger on latency, we conducted a Kruskal-Wallis test, a non-parametric test, to determine whether samples originate from the same distribution. We found that latency is significantly lower on ledger one than ledger two through pairwise comparisons of all latencies recorded from a specific measurement client: t2.Micro ($k=234.67 p<0.025$) and also the t2.medium ($k=317.06; p<0.025$). In both tests, we adjusted the critical value of p to 0.025 using Bonferroni correction. 

\begin{table}[t]
\small
\setlength\tabcolsep{1pt}
\begin{tabular} {| p{4cm} | r | r | r | r | r | r |}
\hline
& \multicolumn{3}{c}{Ledger One} & \multicolumn{3}{|c|}{Ledger Two} \\
\cline{2-7}
Operation & t2.mic & t2.med & Wilcoxon & t2.mic & t2.med & Wilcoxon \\
\hline
\texttt{\detokenize{READ_DID}}                                              & 435 (9) & 406 (7) & Z=38.715& 439 (8) &  410 (9)& Z=38.639 \\
\texttt{\detokenize{READ_CLAIM_DEF}}                                  & 450 (9) & 419 (8) & Z=38.637 & 453 (9) & 423 (10) & Z=38.630  \\
\texttt{\detokenize{READ_REVOC_DEF}}                       & 438 (9) & 410 (8)  & Z=38.556   & 441 (9)  &  416 (8)& Z=38.661  \\
\texttt{\detokenize{READ_REVOC_REG}}                                & 435 (9) & 410 (7) & Z=38.641   & 440 (9) &  413 (8) & Z=38.553 \\
\texttt{\detokenize{READ_REVOC_DELTA}}                   & 47 (16) & 27 (10) & Z=31.248   & 75 (22) & 55 (17) & Z=24.795  \\
\texttt{\detokenize{WRITE_DID}}                                           & 92 (15) & 95 (14) & Z=-7.332   & 162 (9) &  162 (9) &\cellcolor{lightgray}Z=0.564 \\
\texttt{\detokenize{WRITE_REVOC_ADD}}     & 98 (11) &  99 (11) & \cellcolor{lightgray}Z=-2.535   & 166 (12) & 166 (11) & \cellcolor{lightgray} Z=0.324 \\
\texttt{\detokenize{WRITE_REVOC_DEL}}     & 993 (11) & 1000 (11)  & Z=-16.095 & 994 (7) & 1001 (9) & Z=-24.3   \\
\hline
\end{tabular}
\caption{Experiment One: the specification of the measurement client has a significant effect on the latency of read requests, however this pattern is not noticeable for write requests.  Each cell is the median (and interquartile range) of 1000 requests. All Wilcoxon tests have statistical significance at $p<0.003125$, where $p$ is adjusted by Bonferroni correction. Shaded cells are not statistically significant. }
\label{tab:baselineResults} 
\end{table}

\begin{figure}[t]

\includegraphics[width=0.5\textwidth]{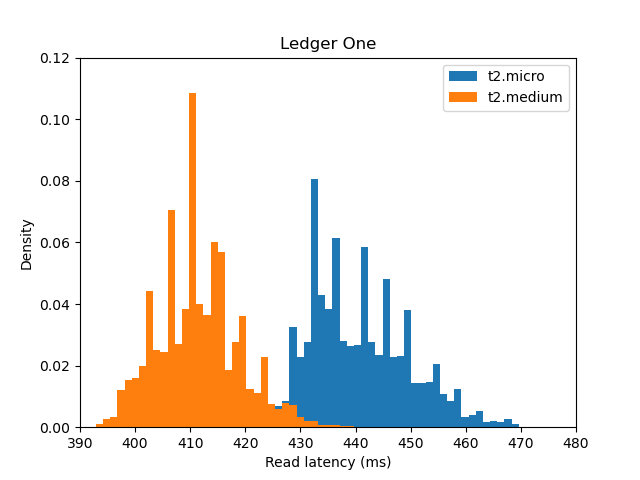}
\includegraphics[width=0.5\textwidth]{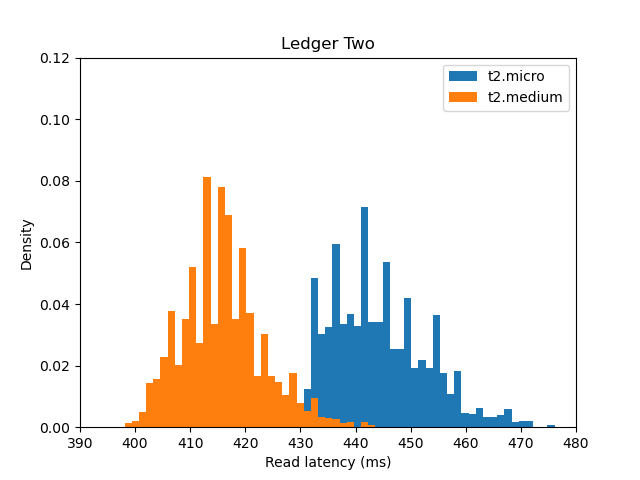}
\includegraphics[width=\linewidth]{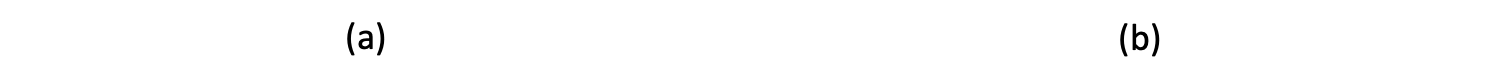}
\includegraphics[width=0.5\textwidth]{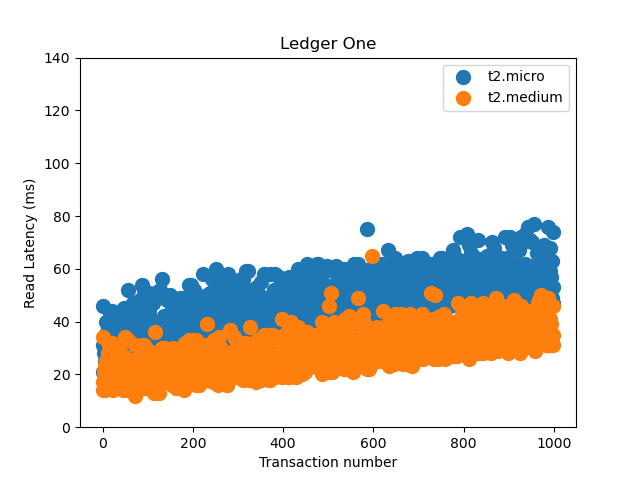}
\includegraphics[width=0.5\textwidth]{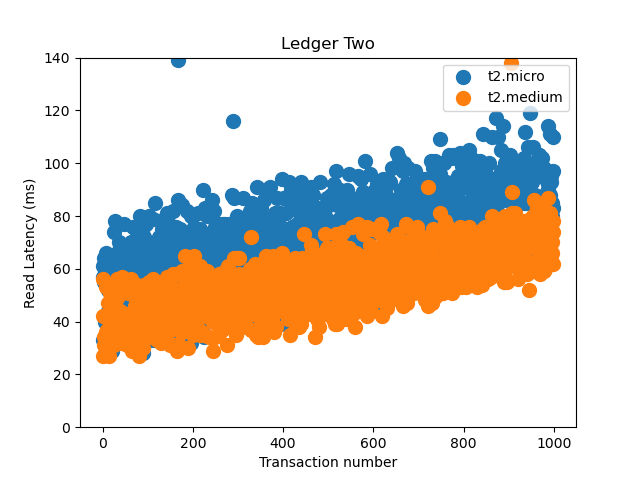}
\includegraphics[width=\linewidth]{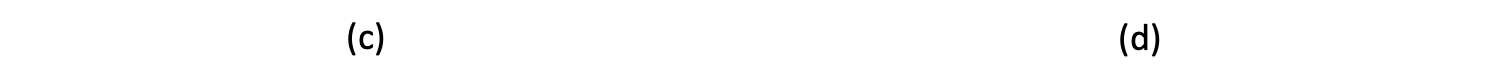}

\caption{Experiment One: the specification of the measurement client impacts read latency. The graphs illustrate the latency of all read operations (excluding READ\_REVOC\_DELTA) on (a) \ledgerone \& (b): \ledgertwo. The latency of the operation READ\_REVOC\_DELTA increases as the number of credentials in an accumulator increases. The observation on (c) \ledgerone; and (d) \ledgertwo.}
\label{fig:readRequests}
\end{figure}

\textbf{Read Requests:} Figure \ref{fig:readRequests} illustrates the read latencies that we recorded. Generally, we observed lower read latency where the measurement client had the more powerful specification, i.e., a t2.medium rather than a t2.micro. 

For ledger one, we found that the median read latency (and interquartile range) is 436 (16) using the t2.micro and 409 (14) using the t2.medium. Indeed, we noted a significant difference between the read latencies recorded by the t2.medium compared to the t2.micro. However, the magnitude of these differences is negligible (see Figure \ref{fig:readRequests}). Therefore, increasing the specification of the Measurement Client from t2.micro to t2.medium can reduce the latency of read requests by around 7\%. We noted a similar trend in the observation of ledger two. The median read latency is 440 (16) from the t2.micro and 413 (13) from the t2.medium. Read requests thus face a performance bottleneck from local computation constraints and ambient network conditions. Note that we did not include the process of message signing by the client in our measurements.

The patterns across all read requests are similar, except for the operation READ\_REVOC\_DELTA. This specific operation yielded smaller latencies than the other read transactions. Furthermore, we noted a strong positive correlation between the transaction number in the experiment and the latency of the operation. The correlation is statistically significant on \ledgerone with the t2.micro ($r=0.65$ $p<0.01$); and t2.medium ($r=0.71$ $p<0.01$); also for \ledgertwo with the t2.micro ($r=0.62$ $p<0.01$) and the t2.medium ($r=0.75$ $p<0.01$). The underlying cause of this pattern is that we add one new credential to a credential accumulator for each experiment round. Therefore the correlation reflects a round-by-round increase in computation cost. In other words, the more credentials registered on the ledger in an accumulator, the more latency to operate on the accumulator. The latency increases by more than 100\% from when the credential accumulator contains a few items to when it is almost full (1000 items). 

\begin{figure}[t]
\includegraphics[width=0.5\textwidth]{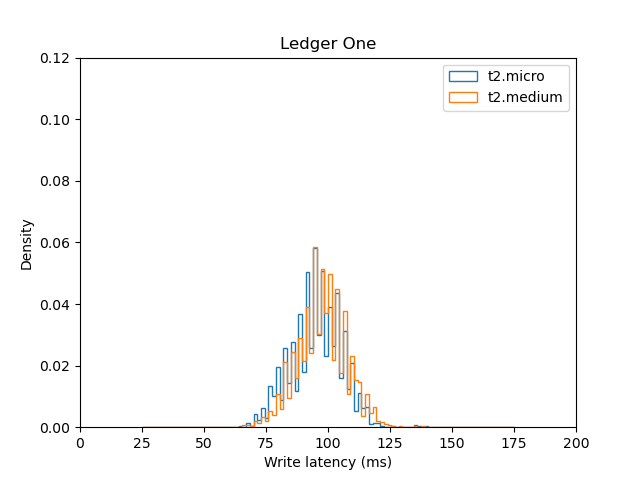}
\includegraphics[width=0.5\textwidth]{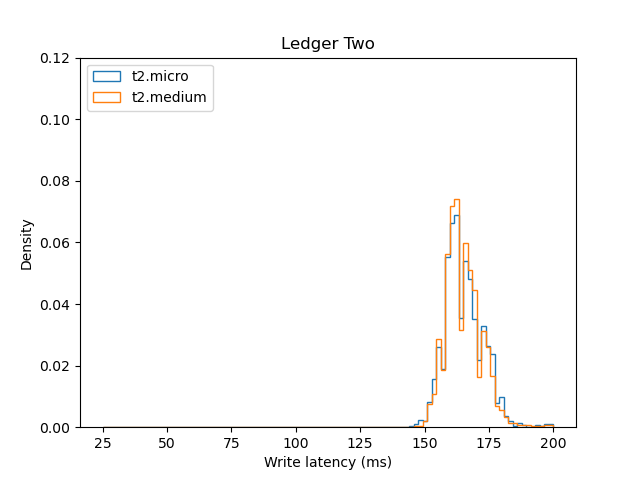}
\includegraphics[width=\linewidth]{figitem1.png}
\caption{Experiment One: the specification of the Measurement Client has little impact on write latency, however the number of ledger nodes exhibits a greater impact. Write latencies (excluding WRITE\_REVOC\_DEL) on (a) \ledgerone; (b) \ledgertwo.}
\label{fig:writeRequests}
\end{figure}

\textbf{Write Requests:} Figure \ref{fig:writeRequests} illustrates the distribution of write latencies. For \ledgerone, the median (and interquartile range) of all write requests is 102 (895) on the t2.micro and 104 (900) on the t2.medium. On \ledgertwo, the median write latency is 170 (828) on the t2.micro, and 169 (835) on the t2.medium. 

Irrespective of the specific ledger and the specification of the measurement client, we measured the operation WRITE\_REVOC\_DEL in a consistently narrow range of 993-1001ms. However, the remaining requests: WRITE\_DID; and WRITE\_REVOC\_ADD, demonstrate increased variation, particularly on \ledgertwo. The specification of the measurement client had a negligible impact on write latency. Nevertheless, as illustrated in Table \ref{tab:baselineResults}, Wilcoxon Signed-Rank Test yielded significant differences between measurements taken on a t2.micro and a t2.medium in three out of six cases. These intermittent effects may be due to a particular dependence of write requests on network conditions subject to fluctuations. A visual inspection of Figure \ref{fig:writeRequests} illustrates minor differences in the medians between the t2.micro and the t2.medium, so we consider that such significant differences may not be important.

\subsection{Experiment Two: Transaction Load Experiment}

\subsubsection{Method}
The experiment is a 2X3 design where the independent variables are ledger (\ledgerone, \ledgertwo) and transaction load (low, medium, high). As before, the dependent variable is transaction latency. Our learnings from Experiment One led us to continue our work using an Amazon EC2 t2.medium machine as the Measurement Client. We select the transaction load parameters of low-medium-high informed by our previous experiences with the stability of Hyperledger Indy: 50-90-100tps for \ledgerone and 10-40-50tps for \ledgertwo. 

The Measurement Client sequentially issues 300 transactions as per the requests from Table \ref{table:transtoproc}. The Measurement Client sends fewer transactions in the second experiment due to the long-expected runtime of an experiment in this context of the same scale. Therefore we measure the latency of 14,400 transactions in total. We configured the \textit{Load Generator} to send transactions to the ledger nodes as quickly as the Hyperledger Indy SDK allows. But we also set a maximum backlog of 1500 un-responded transactions to trigger a pause in the actions of the Load Generator. Since further sending of transactions at this point may trigger a ledger crash.

\begin{table}[t]
\small
\centering
\setlength\tabcolsep{1.5pt}
\begin{tabular} {| l | r  | r  | r  | r | r | r |}
\hline
& \multicolumn{3}{c}{Ledger One} & \multicolumn{3}{|c|}{Ledger Two} \\
\cline{2-7}
Operation & 50tps & 90tps & 100tps & 10tps & 40tps & 50tps \\
\hline
\texttt{\detokenize{READ_DID}}                                              & 414(10) & 684(457) & 1127(19599) & 417(11) &  421(29) & 473(244) \\
\texttt{\detokenize{READ_CRED_DEF}}                                  & 441(16) & 670(557) & 1181(19967) & 437(11) & 471(27) & 492(40)  \\
\texttt{\detokenize{READ_REVOC_DEF}}                       & 416(16) & 455(105)  & 1142(19955)  & 418(11)  &  430(155) & 540(127)  \\
\texttt{\detokenize{READ_REVOC_REG}}                                & 421 (53) & 941 (568) & 1167(19465) & 427(22) &  432(58) & 442(45) \\
\texttt{\detokenize{READ_REVOC_DELTA}}                   & 53(22) & 438(330) & 756(19600)  & 72 (20) & 88(63) & 287(397)  \\
\texttt{\detokenize{WRITE_DID}}                                           & 1558(985) & 3189(888) & 3375 (415)  & 292(45) &  1554(591) & 1439(592) \\
\texttt{\detokenize{WRITE_REVOC_ADD}}     & 1244(414) &  3180(603) & 4233(477) & 892(147) & 1170(378) & 1396(550) \\
\texttt{\detokenize{WRITE_REVOC_DEL}}     & 1145(294) & 4086(725)  & 5723(1619) & 1030(21) & 1187(249) & 1392(655)   \\
\hline
\end{tabular}
\caption{Experiment Two: an increasing ambient transaction load on the ledger nodes affects all operations to different degrees. Each cell is the median (and interquartile range) of 300 operations. }
 \label{tab:experiment2Results} 
\end{table}

\subsubsection{Results}
Table \ref{tab:experiment2Results} illustrates the results of the second experiment. The median latency (and interquartile range) of \ledgerone under a 50tps load is 442 (687); 90tps is 965 (2683), and 100tps is 3603 (6532). For \ledgertwo, the median latency under a 10tps load is 424 (303), a 40tps load is 489 (703), and a 50tps load is 589 (843).

Both ledgers experienced an experimental condition with a transaction load of 50tps. Considering this specific case, we found that latency was significantly lower on \ledgerone than on \ledgertwo ($k=237 p<0.01$).

\textbf{Read Requests:} The overall median read latency (and interquartile range) for \ledgerone at 50tps is 416 (33), at 90tps is 557 (484) and at 100tps is 1123 (19890). For \ledgertwo, the median read latency at 10tps is 420 (23), at 40tps is 426.5 (68) and at 50tps is 474.50 (133).

The read latency on \ledgerone follows a pattern where under the heaviest transaction load, the latency increased by just under a factor of 3 compared to the smallest transaction load. READ\_REVOC\_DELTA appears to be an outlier to this pattern since the median read latency under a 100tps load (756ms) was more than 15 times the median read latency recorded at 50tps (54ms). Though, 100 transactions per second cannot be considered a stable level of traffic for this ledger, as evidenced by the interquartile range for all read requests reaching above 19 seconds.

We observe less elasticity in the latency measurements for \ledgertwo as the transaction load increases. There is a notable increase in the interquartile range as the transaction load increases.

\textbf{Write Requests:} Each write operation exhibits similar variation under varying transaction loads on the network, though the specific ledger clear plays a role too. 

\begin{figure}[t]
\centering
\includegraphics[width=0.49\textwidth]{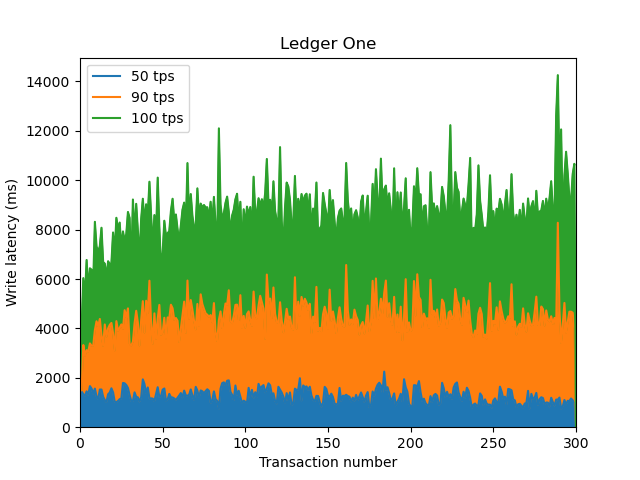}
\includegraphics[width=0.49\textwidth]{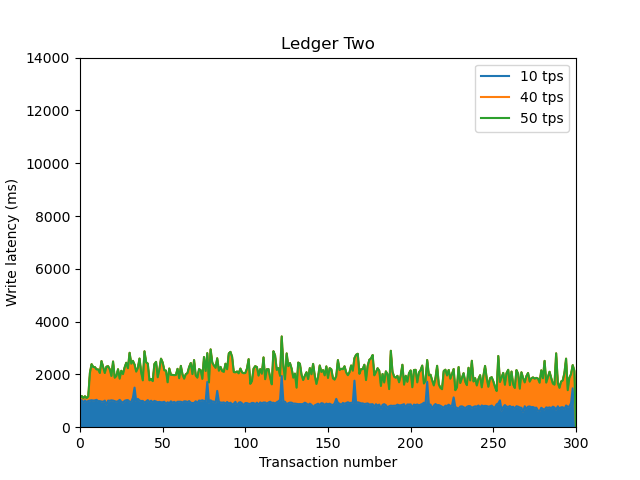}
\includegraphics[width=\linewidth]{figitem1.png}
\caption{Experiment Two: write latency for WRITE\_REVOC\_ADD. Graphs illustrate the measurements across (a) \ledgerone, and (b) \ledgertwo. Latencies are generally lower on \ledgertwo, however \ledgertwo cannot remain stable under a similar transaction load to \ledgerone.}
\label{fig:writeRequests2}
\end{figure}

On \ledgerone, write latency increases by more than 100\% between a transaction load of 50tps and 100tps. The latency of the revocation instruction increases by nearly 5x, and credential creation by 3.4x. On ledger two, the write latency can increase by 7-8 times at 100tps compared to when the network experiences 10tps. What is notable is that \ledgerone shows a greater level of elasticity under a transaction load than \ledgertwo. Figure \ref{fig:writeRequests2} illustrates the write latency for one operation that results in the addition of a credential to the cryptographic accumulator ($WRITE\_REVOC\_ADD$). 

\section{Limitations}
Conducting trustworthy measurement studies is challenging, as well described by Ousterhout~\cite{Ousterhout2018}. While we did our best to mitigate biases in the results of our experiments, there are inevitably limitations that should be taken into account before generalizing or transferring the results to other domains. Our ambition was not to generate an authoritative measure for Hyperledger Indy transaction processing but rather to uncover and document concrete scalability challenges in a software environment relevant to and replicable by others. 

Firstly, we made measurements using code written in \textit{Node.js}, and other programming languages and language-specific dependencies might generate qualitatively different results. Secondly, we can't pinpoint the provenance of every pattern that we uncovered. Still, Hyperledger Indy is open source, and a more determined reader can investigate further given details in our experimental setup in Section \ref{sec:arch}. Finally, Hyperledger Indy is not production-ready software, and so it could be argued that our attention to scalability is somewhat unfair. Given our research is about the blockchain trilemma, we felt that it was still helpful to illustrate the impact of different conceptual design choices (security and decentralization) on an empirical measure (scalability).

\section{Discussion}
\subsection{Increasing the Federation Size Decreases Transaction Throughput and Increases Latency}
Support for large federation sizes is an intuitive requirement to facilitate visions of identity federations comprised of diverse identity providers and relying parties. For example, BankID is a centralized Swedish identity network with ten identity providers that was used 6 billion times by users in 2021 (200 times per second)~\cite{BankID2022}. But similar aspirations face deployment challenges on Hyperledger Indy. While a principle of one-organization-one-node evidences the democratic intent of a decentralized identity scheme, i.e., one node has one vote in the ledger consensus process, large federations of organizations would ultimately reduce the transaction processing capability of the network. The bottleneck is due to the byzantine fault tolerance in Indy's consensus algorithm~\cite{Hyperledger2022} since every node must process all transactions. 

In our first experiment, we found that transaction read latency can be decreased by approximately 7\% by increasing the specification of the measurement client. However, a similar strategy does little to affect write latency. We found In Experiment Two that a ledger with four nodes can sustain a more significant transaction load than a ledger with 11 nodes. This 'sustaining' took the form of ledger nodes processing transactions without the failure of the node itself. We observed relatively low transaction latencies on \ledgertwo, which is a consequence of its reduced ability to function under ambient transaction load, and the tendency for ledger nodes to fail rather than effectively process a backlog of transactions. 

Therefore there might be a point of tension between those who envision large and diverse identity networks, those who expect the structure of one-organization-one-node, and those who advocate byzantine fault tolerance as a consensus method. 

\subsection{Diverse Costs to Process Credential Accumulators}
The case of Hyperledger Indy verifiable credentials (VCs) illustrates the difficulty of making broad-brush conclusions about Trust Registry scalability. Table \ref{tab:baselineResults} shows how read operations for Decentralized Identifiers (DIDs) and other meta-data can be viewed as homogeneous functions in terms of read latency. Except, in the case of VCs, we noticed that the operation to read the accumulated state of the revocation registry between two time periods was ten times faster than the operation to read the revocation registry itself. Also, the write latency to add a new credential to the accumulator is approximately ten times faster than the operation to register the removal of a credential from the accumulator (i.e., to revoke). 

Furthermore, the disturbance we could introduce to those latencies through an ambient transaction load was notable. The write latency for the new credential operation on Ledger One could be stretched by almost four times where the ledger is subject to the heaviest transaction load. In search of a similar pattern on Ledger Two, we observe the credential creation request latency increase by 37\% under the heaviest ambient transaction load. Considering the revocation operation and the heaviest ambient transaction load, the latency can be stretched by 80\% on Ledger One and 27\% on Ledger Two.

Conducting both experiments highlighted that notable computation was spent to locally generate credential accumulators (before they are written to the ledger). The Indy implementation of verifiable credentials features fixed-size cryptographic accumulators (set by parameter). Therefore, in practice, we must estimate the number of credentials that will be needed in advance of creating the accumulator. If the estimate is too low, we must create and manage multiple accumulators in parallel. On the other hand, if we choose to systematically over-estimate the number of credentials that we need, the time to instantiate an Indy credential accumulator increases exponentially with its size. For example, we estimate that a cryptographic accumulator for 1 million credentials would take more than two days to compute based on our experiences. We can thus imagine the time and processes needed to accommodate the 50 million credentials that we proposed as a rule of thumb in section (\ref{sec:deploy}).  

Depending on the context, we may need additional techniques to represent VCs on distributed ledgers like Indy in a privacy-friendly way or different schemes to support the public transparency and revocation of VCs. For example, Sidetrees~\cite{Buchner2021}). 

\subsection{Designing for Non-repudiation}
One interpretation of the design of Hyperledger Indy is that it aims to comprise an authoritative reference point for decentralized identifiers (DIDs) and verifiable credentials (VCs). The desire to provide authoritative statements about DID and VC validity brings to mind Rivest's proposal for a \textit{suicide bureau}~\cite{Rivest1998}: a network that creates real-time positive or negative statements about certificate validity. However, there is one essential item of difference between Rivest's concept and the implementation of Hyperledger Indy: the existence of a \textit{"high-speed reliable network"}. Our experiments demonstrated conditions where the transaction processing of Indy might encounter bottlenecks from a modest number of transactions. The cost of this observation is to the property of non-repudiation, where the signer of a message cannot dispute their own signature.  However, technology design can render aspirations of non-repudiation not meaningful in practice. For example, PGP email encryption is often included (rightly or wrongly) in discussions about identity; however, in practice, it does not provide non-repudiation due to the slow propagation of revocation statements. Thus it can only give the property of authentication~\cite{Roe2010}.  Therefore it remains to be seen whether Indy can underpin an identity network or an authentication network. 

Table \ref{tab:experiment2Results} illustrates how under the heaviest network traffic on our ledger with the fewest nodes, the Interquartile range of one single read of the revocation registry is approximately 19 seconds. One intuitive response to a potential backlog of transactions is to cache reads of the ledger state and queue transactions together (e.g., the revocation registry can be updated locally and its value periodically written to the ledger). Deciding the schedule for this caching is critical; for example, if the revocation state is updated every 24 hours, an attacker may have significant time to abuse a compromised private key. Furthermore, suppose the decentralized identity scheme functions through end-user interaction with a mobile application or cloud storage. In that case, this can also provide plausible reasons why the user is not in control of their private key.

One way to preserve non-repudiation is to re-consider the registration of VCs on the ledger by, e.g., not issuing infinite-term credentials that must be checked for revocation at every usage and instead issue credentials that expire in the short term. For example, the certificate Authority \textit{Let's Encrypt}~\cite{LetsEncrypt} adopts a short certificate lifespan (30 days) on the assumption it is more efficient to re-issue certificates at short time intervals than to issue longer-term credentials and trust the CA revocation infrastructure. 

Another countermeasure is re-imagining how identity providers and relying parties process evidence in an identity network. In the absence of a network with the properties of a \textit{suicide bureau}~\cite{Rivest1998} other, more latent methods of identity verification could become viable. 

\section{Conclusion}
Decentralized identity holds promise as a new approach to digital identity that can improve the privacy of end-users. We considered the challenge to design an appropriate decentralized \textit{Trust Registry} in the face of the challenge to gather context-specific requirements, and the constraints brought by the blockchain trilemma ~\cite{EtherumWiki2022}. We conceptually explored the implications of the blockchain trilemma in the identity context of know-your-customer (KYC) and collated design principles that other researchers can refer to and extend.

In a case study of Hyperledger Indy, we empirically explored one corner of the blockchain trilemma: scalability. We learned that judgments of scalability have multiple facets and made practical observations such as the difficulty to accommodate decentralization through large identity federations, and the deployment challenges facing the credentials-on-the-ledger model. Finally, we considered the existence of an implicit security design goal of non-repudiation, where the signer of a message cannot dispute their own signature. A property that is driven by the potential of asymmetric cryptography but that is difficult to achieve in a system deployment. It is an open question whether non-repudiation can be achieved through a technology with transaction processing limitations. 

Our research highlights the challenge to design contextually useful Trust Registry technology subject to the blockchain trilemma.
 
%\setcounter{tocdepth}{1}
%\listoftodos
\bibliography{sigproc}
\bibliographystyle{plain}

\end{document}